\documentclass[twocolumn,superscriptaddress, amsmath,amssymb]{revtex4-1}
\usepackage{amssymb}
\usepackage{amsmath}
\usepackage{graphicx, subfigure}
\usepackage{color}
\usepackage{ulem}
\usepackage{graphicx}
\usepackage{dcolumn}
\usepackage{bm}

\def\eps{\varepsilon}

\date{\today}

\begin{document}
\title{Asymmetric coupling and dispersion of surface-plasmon-polariton waves on a periodically patterned anisotropic metal film}
\author{Jhuma Dutta}\author{S. Anantha Ramakrishna}
\affiliation{Department of Physics, Indian Institute of Technology Kanpur, Kanpur 208016, India}
\author{Akhlesh Lakhtakia}
\affiliation{Department of Engineering Science and Mechanics, The Pennsylvania State University, University Park, Pennsylvania 16802, USA}
\date{\today}

\begin{abstract}
The morphology of a columnar thin film (CTF) of silver renders it an effectively biaxially  anisotropic continuum.  CTFs of silver  deposited on one-dimensional gratings of photoresist showed strong blazing action and  asymmetrically coupled optical radiation to surface plasmon-polariton (SPP) waves propagating only along one direction supported by either the CTF/photoresist or the CTF/air interfaces. Homogenization of the CTFs using the Bruggeman formalism revealed them to display hyperbolic dispersion, and the dispersion of SPP waves was adequately described thereby.

\end{abstract}

\maketitle

\section{Introduction}
Surface plasmon-polariton (SPP) waves guided by the  interface of a metal and a dielectric medium have been projected to carry information in future miniaturized all-optical plasmonic chips~\cite{science_ozbay}. Control over the dispersion of SPP waves  and their coupling to light are critical for this and other applications. While SPP waves supported by the interface of  a metal and a dielectric medium, both isotropic and homogeneous, are well known for several decades~\cite{maier_book}, interest has arisen in the case of the partnering dielectric medium being periodically nonhomogeneous normal to the interface~\cite{FLjosab2010}, which allow the possibility of multiple SPP waves that will enable  extremely sensitive multi-analyte sensors~\cite{lakhtakia_sci_rep2013}.  

An  important aspect is the excitation of SPP waves by incident light, which is usually mediated by a dielectric or a grating coupler~\cite{maier_book}. Coupling of light solely into SPP waves propagating along a single direction is an important requirement in many instances, which can be achieved by arrangements such as oblique  incidence on a grating~\cite{laluet_OE2007} or by interference effects made possible by appropriately placed nanoscatterers~\cite{Bozhevolnyi_RPP2013}. Slanted metal gratings were predicted to couple light preferentially into SPP waves propagating in a specific direction~\cite{bonod_OE2007} and slanted sinusoidal gratings and binary blazed plasmonic gratings have been similarly used~\cite{bai_PRB2009}.

Control of propagation for applications such as subwavelength focusing of  SPP waves~\cite{kadic_ACSnano2011} requires not only anisotropic dielectric materials, but anisotropic metals as well. The CTF of a metal, deposited by an obliquely incident collimated metal vapor flux on a substrate \cite{lakhtakia_book}, is highly anisotropic, as has been demonstrated for CTFs of plasmonic metals such as silver and  gold~\cite{shalabney_PNFA2009} as well as magnetic materials like cobalt~\cite{schubert_APL2012}. Periodically patterned CTFs of inorganic dielectric materials such as CaF$_2$ were recently shown to function well as blazed diffraction gratings~\cite{jhuma_APL2013}. Hence, it is attractive to investigate the plasmonic properties of periodically patterned CTFs made of plasmonic metals. 

\begin{figure}
\centering
\includegraphics[width=4.2cm]{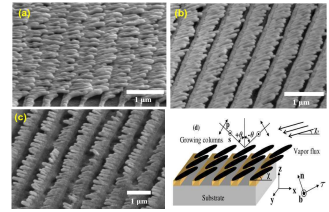}
\caption{\label{fig_FESEM} FESEM images of  periodically patterned CTFs of silver columns with (a) $d=480$~nm, (b) $d=580$~nm, and (d) $d=770$~nm. (d) Schematic showing the columnar morphology and the incident light, along with the principal axes $(\hat{\tau},\hat{b},\hat{n})$ of the CTF while $(x,y,z)$ are the Cartesian coordinates.}
\end{figure}

In this paper, lithographically patterned submicron gratings of silver CTFs illustrated in Fig.~\ref{fig_FESEM} with slanted columns have been shown to provide for a strong unidirectional coupling to SPP waves.  Experiments on the dispersion of  SPP waves have revealed that the silver CTF is an effectively anisotropic metal with hyperbolic dispersion~\cite{smith_PRL2003}. The columnar orientation and the porosity of the CTF completely define the effective permittivity tensor, and thereby provide a route to generate any desired characteristics for the hyperbolic medium. Due to the columnar morphology, not only incident $p$-polarized light but also incident $s$-polarized light indirectly couples to the SPP waves via internally scattered waves.

The plan of this paper is as follows. Section~\ref{emm} provides a description of the experimental methods and
materials used by us. Section~\ref{theory} presents the theory of homogenization of the metal CTF into an effectively biaxially  anisotropic continuum as well as the derivation of the  wavenumber of the SPP wave guided by the planar interface of the homogenized metallic STF and an isotropic dielectric material, when the direction of propagation of the SPP wave lies wholly in the morphologically significant plane of the CTF. Experimental and theoretical results are presented and discussed in Sec.~\ref{rd}. Concluding remarks are provided in Sec.~\ref{conc}.

\section{Experimental Methods and Materials}\label{emm}
Periodically patterned CTFs were fabricated by the deposition of a collimated flux of silver vapor directed obliquely towards 1-D gratings of photoresist (ma-P 1205, Micro-resist Technology) in a vacuum chamber. Several gratings  were made by laser interference lithography using a He-Cd laser (442~nm, 30~mW, Kimmon, Japan) on glass substrates that had been first cleaned and then spin coated with a photoresist. Atomic force microscopy (XE70, Park Systems, S. Korea) was used to determine the  period $d$ as 480, 580, 680, and 770~nm for different gratings. All gratings  had an approximately 50-50 duty cycle and a depth modulation of 150~nm. Silver vapor was directed at an angle $\chi_v=4^\circ$ with respect to the mean plane of the grating. When the collimated vapor flux thus arrives  very obliquely at the photoresist grating, only the ridges  are exposed to the vapor while the valleys are shadowed out. This results in the growth of silver columns only on the ridges. Deposition took place on gratings with 
different periodicities simultaneously to ensure identical deposition conditions. The deposition rate was maintained at $5-7$~\AA~s$^{-1}$ with the help of a quartz crystal monitor.  Base pressure within the  chamber was $4\times10^{-6}$ mbar, which rose to about $1\times 10^{-5}$ mbar during deposition.  CTFs of about 500-nm nominal thickness were grown in this manner. 

The PP-CTFs were imaged using  Field-effect scanning electron microscopy (FESEM) (BP 40 SUPRA Carl Zeiss instruments). A 5~nm layer of gold was deposited at normal incidence to reduce charging effects.

Diffraction efficiencies of the periodically patterned CTFs were measured using diode lasers of wavelengths 442, 532 and 633 nm. A Glan--Thompson polarizer was used to linearly polarize the incident light (either $p$ or $s$). A silicon photodiode placed on a goniometric stage is used to measure the laser intensity at different angles for the zeroth and higher diffraction orders after passing through the samples. Neutral density filters are used to attenuate the beam to avoid saturation of the photodiode. 

Angle-resolved transmittance spectra of the periodically patterned CTFs were determined using a collimated beam from a 100-W tungsten-halogen lamp and a fiber spectrometer (USB4000+, Ocean Optics) with a resolution of 0.5~nm in the 450-1000~nm wavelength range. A Glan--Thompson polarizer was used to linearly polarize (either $p$ or $s$) the incident light.  The polarization of the incident light  and the angle of incidence $\theta$ are indicated on the schematic in Fig.~\ref{fig_FESEM}(d). The sample was placed on a motorized rotation stage and the transmission spectra were obtained at an angular rotation by $1.8^\circ$.

\section{Theory of SPP-wave propagation}\label{theory}

\subsection{Relative-permittivity model for the CTF}

The effective relative permittivity tensor $\tilde{{\eps}}$ of a CTF can be written as~\cite{lakhtakia_book}
\begin{equation}
\tilde{{\eps}} \equiv  \left( \begin{array}{ccc}
{\eps}_{xx} & {\eps}_{xy} & {\eps}_{xz}  \\
{\eps}_{yx} & {\eps}_{yy} & {\eps}_{yz} \\
{\eps}_{zx} & {\eps}_{zy} & {\eps}_{zz} \end{array}\right)
= {\cal S}_y \cdot
{\rm Diag}\left[{\eps}_{b},{\eps}_{c},{\eps}_{a}\right]\cdot
{\cal S}_y^T\,,
\label{eps-CTF1}
\end{equation}
where the  matrix  
\begin{equation}
{\cal S}_y = \left( \begin{array}{ccc}
\cos\chi & 0 & -\sin\chi\\
0 & 1 & 0 \\
\sin\chi & 0 & \cos\chi \end{array} \right)\,,
\end{equation}
indicates a rotation about the $y$ axis.
Whereas ${\eps}_{b}$, ${\eps}_{c}$, and ${\eps}_{a}$ are the eigenvalues of $\tilde{{\eps}}$,
$\hat{\tau}=\hat{x}\cos\chi+\hat{z}\sin\chi$, $\hat{b}=-\hat{y}$, and $\hat{n}=-\hat{x}\sin\chi+\hat{z}\cos\chi$ are the corresponding eigenvectors of $\tilde{{\eps}}$.

The eigenvalues of $\tilde{{\eps}}$ can be estimated using a Bruggeman formalism wherein the bulk metal and the voids are supposed to be distributed as prolate ellipsoids~\cite{lakhtakia_book}, so long as the porosity (i.e., the void volume fraction) $f_v$ is sufficiently large \cite{MLoc2004}. That being true for all the CTFs investigated by us, the Bruggeman formalism requires the solution of the three coupled equations expressed compactly
as 
\begin{eqnarray}
\nonumber
&&
\left(1-f_v\right) \left({\eps}_m{\cal I} -\tilde{{\eps}}\right)\cdot
\left[{\cal I}+{i\omega{\eps}_0}\uline{\uline{{\cal D}}}^{(m)}\cdot
\left({\eps}_m{\cal I} -\tilde{{\eps}}\right)\right]^{-1}
\\
&&
+f_v \left({\eps}_v{\cal I} -\tilde{{\eps}}\right)\cdot
\left[{\cal I}+{i\omega{\eps}_0}\uline{\uline{{\cal D}}}^{(v)}\cdot
\left({\eps}_v{\cal I} -\tilde{{\eps}}\right)\right]^{-1}=0\,.
\label{eq6}
\end{eqnarray}
Here, ${\eps}_0$ is the permittivity of free space; ${\eps}_m$ and ${\eps}_v=1$ are the relative permittivity scalars  of the bulk metal and the voids, respectively;
${\cal I}$ is the idempotent;  the  depolarization tensors
\begin{widetext}
\begin{equation}
\uline{\uline{{\cal D}}}^{(m,v)} = \frac{2}{i\pi\omega{\eps}_0}\int_{\varphi=0}^{\frac{\pi}{2}}\int_{\vartheta=0}^{\frac{\pi}{2}}\frac{(\sin\vartheta\cos\varphi)^2\, \hat{n}\hat{n} 
+(\sin\vartheta\sin\varphi/\gamma_{b}^{(m,v)})^2\, \hat{b} \hat{b} 
+(\cos\vartheta/\gamma_{\tau}^{(m,v)})^2\, \hat{\tau}\hat{\tau}}{(\sin\vartheta\cos\varphi)^2{\eps}_{a}+(\sin\vartheta\sin\varphi/\gamma_{b}^{(m,v)})^2{\eps}_{c}
+(\cos\vartheta/\gamma_{\tau}^{(m,v)})^2{\eps}_{b}}\,\sin\vartheta\,d\vartheta\,d\varphi\,;
\end{equation}
\end{widetext}
and
 $\gamma_{\tau,b}^{(m,v)}$ are the shape parameters of the ellipsoids \cite{lakhtakia_book}. 
 
 The parameter $\chi$ can be determined from FESEM images.  The parameters $\gamma_{\tau,b}^{(m,v)}$ and $f_v$ have to be chosen by comparing theoretical predictions of optical response characteristics against their experimental counterparts \cite{SLH}.

\subsection{SPP wavenumber}

Suppose  that the metallic CTF occupies the half space $z>0$ and the isotropic dielectric material occupies the half space $z<0$. Thus,  SPP-wave propagation is guided by the plane $z=0$.  As the $xz$ plane is the morphologically significant plane of the CTF, any SPP wave propagating along the $x$ axis in the $xy$ plane can be either $s$ polarized (i.e., $E_x=E_z=0$ and $H_y=0$) or $p$ polarized (i.e., $E_y=0$ and $H_x=H_z=0$).

The $p$-polarized electromagnetic fields in the homogenized metallic CTF ($z > 0$) are given by
\begin{equation}
\left.\begin{array}{l}
\vec{E}_m = E_{xm} \left(\hat{x} + \frac{ik_xk_{zm}+k_0^2{\eps}_{zx}}{k_x^2 - k_0^2{\eps}_{zz}}\hat{z}\right) e^{i(k_xx+ik_{zm}z)}\\
\vec{H}_m = -E_{xm} \left(\hat{y}\omega{\eps}_0\frac{{\eps}_{zx}k_x+ik_{zm}{\eps}_{zz}}{k_x^2 - k_0^2{\eps}_{zz}}\right) e^{i(k_xx+ik_{zm}z)}
\end{array}
\right\}\,,
\label{CTF}
\end{equation}
and in the isotropic dielectric material ($z < 0$)   by
\begin{equation}
\left.\begin{array}{l}
\vec{E}_d = E_{xd} \left(\hat{x} - \frac{ik_x}{k_{zd}}\hat{z}\right)e^{i(k_xx-ik_{zd}z)}\\
\vec{H}_d = E_{xd} \left(\hat{y}\omega{\eps}_0\frac{i{\eps}_d}{k_{zd}}\right)e^{i(k_xx-ik_{zd}z)}
\end{array}
\right\}\,,
\label{diel}
\end{equation}
where $k_0$ is the free-space wavenumber.

Substitution of Eqs.~(\ref{CTF}) into the Maxwell curl equations  for the CTF leads to the relationship
\begin{equation}
k_x^2{\eps}_{xx}+2ik_xk_{zm}{\eps}_{xz}-k_{zm}^2{\eps}_{zz}-k_0^2{\eps}_a{\eps}_b=0\,
\end{equation}
between $k_x$ and $k_{zm}$.  Two values of $k_{zm}$ exist for every $k_x$; we must choose that value
 which satisfies the constraint $\mathrm{Re}\left(k_{zm}\right)>0$ in order to ensure decay of fields as
 $z\to\infty$. Likewise, the dispersion equation
 \begin{equation}
 k_x^2-k_{zd}^2-k_0^2{\eps}_{d}=0\,,
 \end{equation}
 follows from the substitution of Eqs.~(\ref{diel})  into the Maxwell curl equations  for the isotropic dielectric material.
  Two values of $k_{zd}$ exist for every $k_x$; we must choose that value
 which satisfies the constraint $\mathrm{Re}\left(k_{zd}\right)>0$ in order to ensure decay of fields as
 $z\to-\infty$. Enforcing the continuity of $E_x$ and $H_y$ across the interface $z=0$ yields the dispersion
 equation
 \begin{equation}
k_x^2{\eps}_d-ik_{zd}k_x{\eps}_{zx}+(k_{zd}k_{zm}-k_0^2{\eps}_d){\eps}_{zz}=0\,,
 \label{predict}
\end{equation}
for SPP-wave propagation. The solution $k_x$ of Eq.~(\ref{predict}) is the wavenumber  $k_{spp}$ of the
$p$-polarized SPP wave propagating along the $x$ axis.

\section{Results and Discussion}\label{rd}

\subsection{Morphology}
FESEM images of the samples (see Fig.~\ref{fig_FESEM}) revealed the growth of well-oriented long nanocolumns of silver on the ridges with an average length of $500\pm 10$ nm, a diameter of about $80\pm 10$ nm, and a tilt angle $\chi\simeq 8^\circ \pm 2^\circ$ with respect to the mean plane of the grating. These statistics were obtained by averaging over 25 
nanocolumns in each FESEM image.

\subsection{Asymmetric diffraction}
Overall, the fabricated samples were fairly smooth, and scattered very little light from the CTF regions outside the grating area ($\sim 20$~$\mathrm{mm}^2$), while exhibiting strong diffraction orders in the reflection from the CTF-coated grating regions.

\begin{figure*}
\includegraphics[width=10cm]{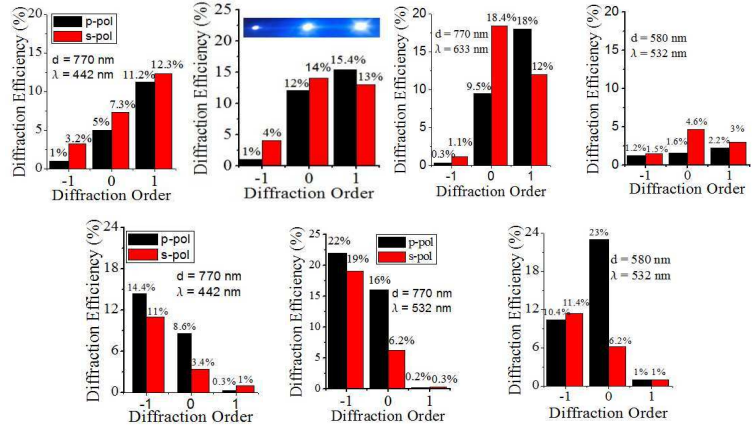}
\caption{Top row: From left to right, the first three bar diagrams present the measured diffraction efficiencies in the transmission mode for $d = 770$~nm and $\lambda\in\left\{442, 532, 633\right\}$~nm for incident $p$-polarized and incident $s$-polarized light, and  the fourth diagram is for $d = 580$~nm and $\lambda = 532$~nm. Bottom row: Measured diffraction efficiencies in the reflection mode for $d \in\left\{770,580\right\}$~nm and $\lambda\in\left\{442, 532\right\}$~nm for incident $p$-polarized and incident $s$-polarized light. The inset is a photograph showing asymmetric diffraction for $\lambda$ = 442 nm. \label{fig_diff_eff}}
\end{figure*}

Highly asymmetric diffraction patterns were exhibited by the periodically patterned silver CTFs,  similarly to the blazed diffraction by periodically patterned dielectric CTFs~\cite{jhuma_APL2013}. The diffraction efficiencies (i.e., the ratio of the diffracted intensity to the incident intensity) measured for the various orders are tabulated in Fig.~\ref{fig_diff_eff} for incident $p$ and $s$-polarized light, when the period $d \in\left\{770,580\right\}$~nm and the wavelength $\lambda\in\left\{442, 532, 633\right\}$~nm in the transmission mode (top row) and in the reflection mode (bottom row). The asymmetry between the $n=+1$ and $n=-1$ orders are highly pronounced for incident $p$-polarized light than for incident $s$-polarized light. One photograph of the diffraction pattern was taken using $\lambda=442$~nm wavelength is shown in the inset of Fig.~\ref{fig_diff_eff} which shows asymmetric diffraction. 

Diffraction efficiencies are functions of the height, period, and the duty cycle of a slanted  grating~\cite{bonod_OE2007,bai_PRB2009}. Although full-scale three-dimensional computations (see later) will be required for optimization, the  periodically patterned silver CTFs clearly show potential 
for uni-directional coupling of light to SPP waves.

\subsection{Coupling of $p$-polarized light to SPP waves}

\begin{figure*}
\includegraphics[width=12cm]{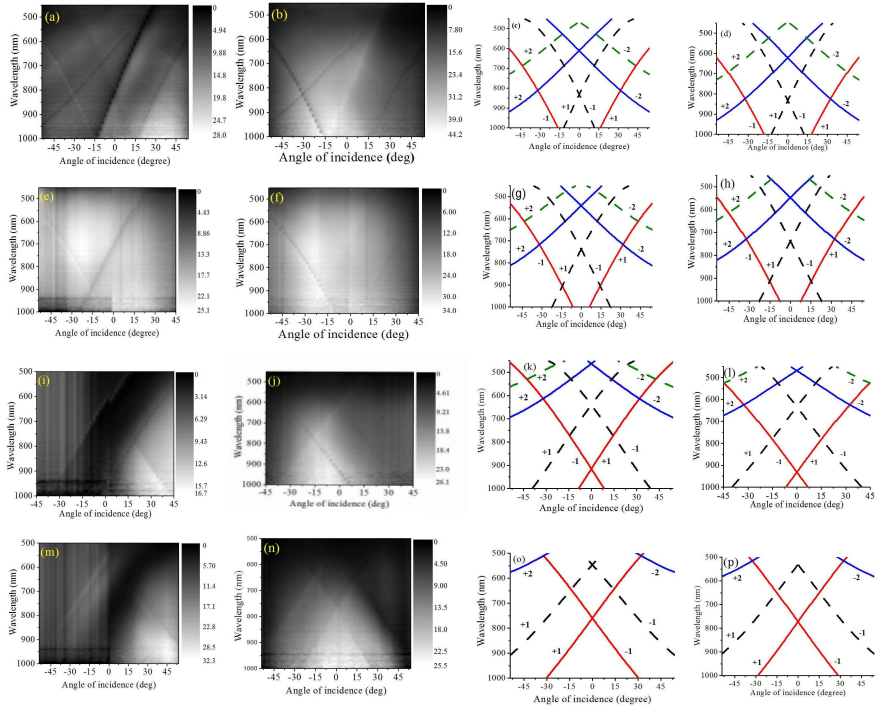}
\caption {Angle-resolved transmittance spectra excited when the incident  light is $p$ polarized  and the period $d=$ (a) 770~nm, (e) 680~nm, (i) 580~nm, and (m) 480 nm; angle-resolved transmittance spectra excited when the incident  light is $s$ polarized  and the period $d=$ (b) 770~nm, (f) 680~nm, (j) 580~nm, and (n) 480~nm; theoretical plots for the dispersion of SPP waves obtained  using $\tilde{{\eps}}$ predicted by the Bruggeman model with ellipsoidal voids ($f_v=0.88$, $\gamma_{\tau}^{(m)}=7$, $\gamma_{b}^{(m)}=1.2$, $\gamma_{\tau}^{(v)}=10$, and $\gamma_{b}^{(v)}=1$) when the period $d=$ (c) 770~nm, (g) 680~nm, (k) 580~nm, and (o) 480~nm; and  theoretical plots for the dispersion of SPP waves obtained using $\tilde{{\eps}}$ predicted by the Bruggeman model with spherical voids ($f_v=0.91$, $\gamma_{\tau}^{(m)}=15$, $\gamma_{b}^{(m)}=1.5$, $\gamma_{\tau}^{(v)}=1$,  and $\gamma_{b}^{(v)}=1$) when the period $d=$ (d) 770~nm, (h) 680~nm, (l) 580~nm, and (p) 480~nm. The theoretical 
predictions for the CTF/photoresist interface are solid lines, while those for the CTF/air interface are dashed lines. The orders $n = \pm1$ and $\pm2$ for  Bragg scattering are indicated. \label{fig_dispersions}}
\end{figure*}

Evidence of coupling to SPP waves is available in the measured angle-resolved transmission spectra
presented in Fig.~\ref{fig_dispersions}.   Given the large thickness of the silver CTF, SPP waves localized at the  CTF/photoresist interface will not be detuned due to scattering from the CTF/air interface   and vice versa~\cite{maier_book}. When $\theta$ satisfies the condition 
\begin{equation}
\vec{k}_{spp}  = k_0 \sin\theta\,\hat{x}   +  n(2\pi/d)\hat{x} \,
\label{eq1}
\end{equation}
for resonant excitation of an SPP wave,
where $\vec{k}_{spp}=k_{spp}\;\hat{x}$  and $n\ne0$ is either a positive or a negative integer, the incident light couples to the SPP wave and there is a strong attenuation of the transmitted field.  The transmittance minima trace out the dispersions of the SPP waves. 

Figures~\ref{fig_dispersions}(a), (e), (i),  and (m) show the angle-resolved transmittance spectra obtained with $p$-polarized light for four values of $d$. If the CTF were replaced by a dense metal film, the coupling to the SPP waves would be symmetric with respect to the sign of $\theta$. Instead,  the four figures evince  a strong asymmetry with respect to  positive and negative $\theta$, and only one of the two branches is present in most cases. Momentum transfer from the grating is possible in this case only for $n>0$, as will become clear in the sequel. While a gross angularly asymmetric transmittance with respect to the normal by slanted metallic CTFs is well known~\cite{gbsmith_AO1990}  and is also evident from  Fig.~\ref{fig_dispersions}, it had not been hitherto realized that slanted metallic CTFs deposited on gratings can be useful for unidirectional coupling to SPP waves.

 In order to compare theory with experiment, the formulation presented in Sec.~\ref{theory} was employed.
 The Drude model was used for the relative permittivity of silver as ${\eps}_m(\omega) = 5.7 - \omega_p^2/[\omega(\omega + i\gamma_p)]$, where the angular frequency $\omega$ is in units of eV, $\omega_p = 9.2$~eV, and $\gamma_p = 0.021$~eV. The parameters for homogenization with ellipsoidal voids ($f_v=0.88$, $\gamma_{\tau}^{(m)}=7$, $\gamma_{b}^{(m)}=1.2$, $\gamma_{\tau}^{(v)}=10$, and $\gamma_{b}^{(v)}=1$) and spherical voids ($f_v=0.91$, $\gamma_{\tau}^{(m)}=15$, $\gamma_{b}^{(m)}=1.5$, $\gamma_{\tau}^{(v)}=1$, and $\gamma_{b}^{(v)}=1$) were used to first estimate $\tilde{{\eps}}$ using Eq.~(\ref{eq6}) and then obtain $k_{spp}$ using Eq.~(\ref{predict}). 
 
\begin{figure*}
\includegraphics[width=12cm]{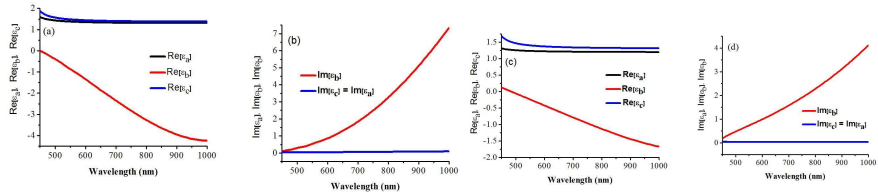}
\caption{(a) Real and (b) imaginary parts of ${\eps}_{a,b,c}$ predicted as functions of the wavelength by the Bruggeman model with ellipsoidal voids ($f_v=0.88$, $\gamma_{\tau}^{(m)}=7$, $\gamma_{b}^{(m)}=1.2$, $\gamma_{\tau}^{(v)}=10$, and $\gamma_{b}^{(v)}=1$); (c) real and (b) imaginary parts of ${\eps}_{a,b,c}$ predicted as functions of the wavelength by the Bruggeman model with spherical voids ($f_v=0.91$, $\gamma_{\tau}^{(m)}=15$, $\gamma_{b}^{(m)}=1.5$, $\gamma_{\tau}^{(v)}=1$,  and $\gamma_{b}^{(v)}=1$). \label{fig_epsilon_comps}}
\end{figure*}

 Figure~\ref{fig_epsilon_comps} shows the variations of ${\eps}_{a,b,c}$ with wavelength. Clearly, the real part of the projection ${\eps}_b$ of $\tilde{{\eps}}$ on the columnar axis $\hat{\tau}$ is negative over the visible and near-infrared frequencies, while ${\rm Re}[{\eps}_{a,c}]>0$. Hence, the metal CTF is a biaxial medium with an indefinite relative permittivity
 ternsor. Also, the homogenization model predicts a large imaginary part of ${\eps}_b$.

The predicted  occurrences of SPP waves localized to the interface of this hyperbolic medium with the photoresist (${\eps}_d=2.6$)  are shown in Figs.~\ref{fig_dispersions}(c), (g), (k), and (o) (for ellipsoid voids)   with $d = 770$, $680$, $580$, and $480$~nm, respectively. Likewise,
the predicted  occurrences of SPP waves localized to the interface of the same hyperbolic medium with   air (${\eps}_d=1$) are shown in   
Figs.~\ref{fig_dispersions}(d), (h), (l), and (p) (for spherical voids) with $d = 770$, $680$, $580$, and $480$~nm, respectively. 
Just a single value of $f_v$  ($= 0.88$) sufficed to predict the dispersions accurately for all three values of the period $d$, when compared with the experimentally obtained Figs.~\ref{fig_dispersions}(a), (e), (i),  and (m).
The FESEM images also indicate $f_v \simeq 0.9$ for all three silver CTFs.

\begin{figure*}
\includegraphics[width=12cm]{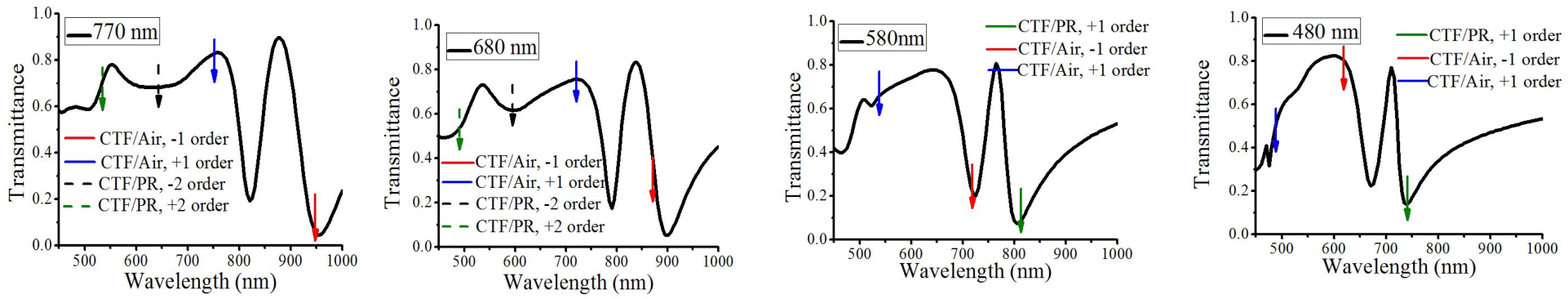}
\caption{Simulated transmittance spectra through the structure with period $d\in\left\{ 770, 680, 580, 480\right\}$~nm when $s$-polarized light is incident at $\theta=5^\circ$. Red and blue solid-line arrows correspond to SPP waves guided by the CTF/air interface with $n=-1$ and $n=+1$ orders, respectively,  black and green solid-line arrows correspond to SPP waves guided by the CTF/photoresist interface with $n=-1$ and $n=+1$ orders, respectively; and black and green dotted-line arrows correspond to SPP waves guided by CTF/photoresist interface with $n=-2$ and $n=+2$ orders, respectively.\label{transmission_spectra}
}
\end{figure*}

 \begin{figure*}
\includegraphics[width=5.9cm]{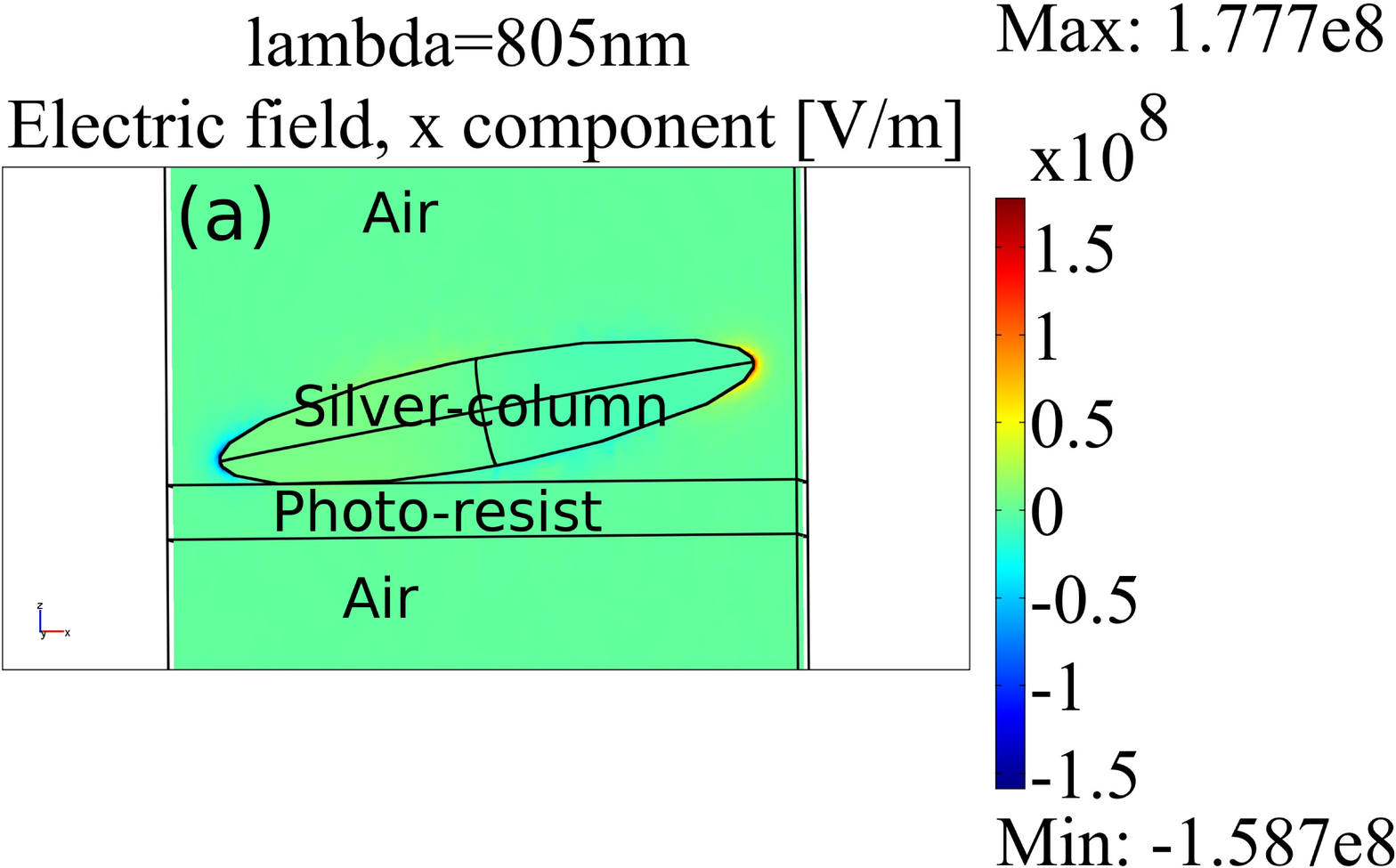}
\includegraphics[width=5.9cm]{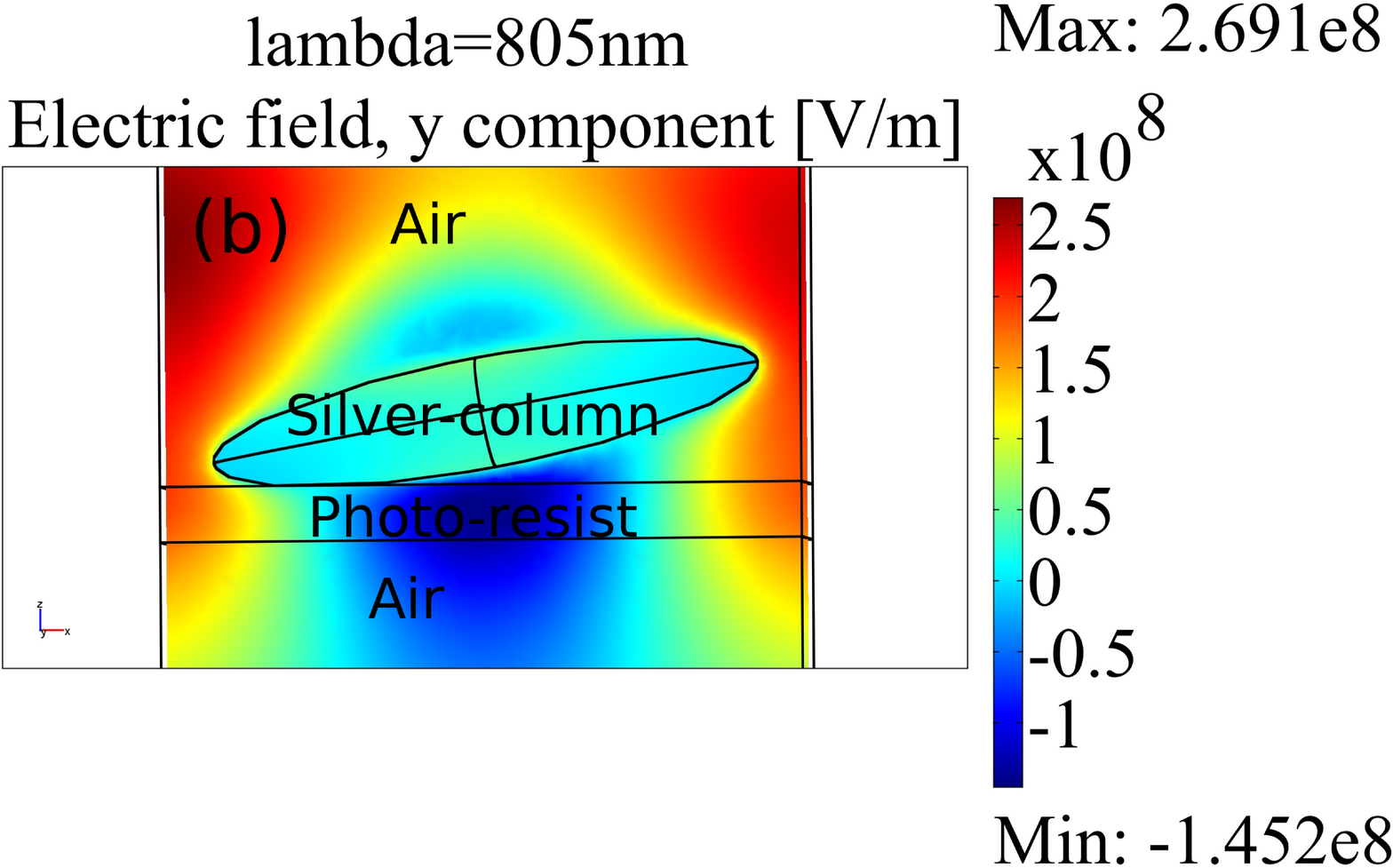}
\includegraphics[width=5.9cm]{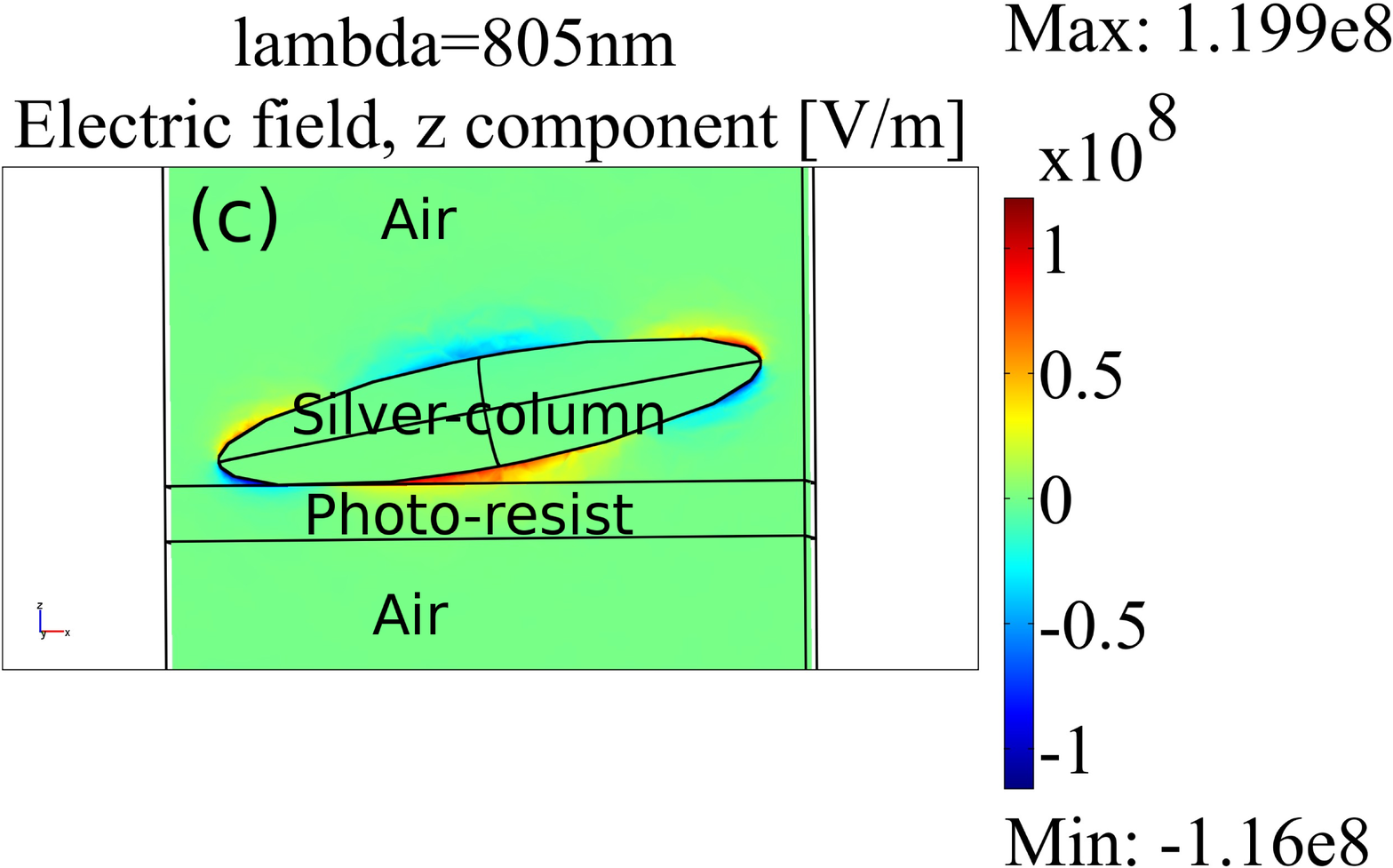}
\caption{Simulated distributions of (a) $E_x$, (b) $E_y$, and (c) $E_z$  in the structure with period $d = 580$~nm illuminated by $s$-polarized light of free-space wavelength 805~nm incident at $\theta=5^\circ$. The incident plane wave of the form $\exp(-i\omega t)$ was assumed and the shown fields are in phase with the incident wave. Lengths of major and two minor axes of the ellipsoids are 500~nm and 90~nm, respectively.
 \label{Electric_fields}}
\end{figure*}

\subsection{Coupling of $s$-polarized light to SPP waves}

A theoretical treatment of the metallic CTF as a homogeneous continuum  indicates that $s$-polarized incident light cannot couple to an SPP wave in the present situation. Yet, in Figs.~\ref{fig_dispersions}(b), (f), (j), and (n) dispersive features indicative of that coupling are evident in the angle-resolved transmittance spectra obtained with $s$-polarized light. The 
$s$-polarized incident light could couple to the SPP waves due to scattering arising from either the columnar structure of the CTF or from the surface roughness that is visible in the FESEM images. To verify whether the columnar structure itself is responsible for this effect, we performed numerical  simulations considering only the columnar structure with no roughness. 

The simulations were performed using the finite-element method implemented with COMSOL Multiphysics software (Version 3.5a, 3-dimensional harmonic propagation mode in RF module) to validate the coupling of $s$-polarized light to SPP waves  even in the absence of surface roughness. A periodic array of silver prolate ellipsoids with their major axes oriented along the nanocolumnar axis (i.e., $\hat{\tau}$) was considered as the CTF. The ellipsoids were placed on a dielectric substrate of refractive index $1.65$. Perfect-electrically-conducting   boundary conditions were applied along the $y$ direction to the faces perpendicular to $\bf E$, and periodic boundary conditions consistent with the Floquet theory were applied along the $x$ directions on the faces parallel to $\bf E$. The unit cell size was $d$ $\times$ 110~nm $\times$ 700~nm. Perfectly matched layers were used above and below the periodic array to prevent multiple reflections of the incident wave. Transmittance spectra were calculated by 
integrating the power flow on the planes 
immediately below and above the array of ellipsoids.

The transmission through a periodic array of appropriately oriented silver ellipsoids revealed strong spectral minima that approximately coincide with the measured transmittance minima for all the grating periods $d$ considered for this paper. We show plots of the simulated transmittance spectra for  $d\in\left\{ 770, 680, 580, 480\right\}$~nm in Fig.~\ref{transmission_spectra}. The transmittance dips in the plots should correspond to coupling with SPP waves. 

Figure~\ref{Electric_fields} shows the magnitude of the normal component ($E_z$) of the electric field excited at resonance by the $s$-polarized incident light (which does not have $x$- and $z$-directed components). Very significantly, large localized $E_x$ and $E_z$  are excited of similar magnitudes as  $E_y$.

Thus, it becomes clear that the scattered fields, particularly the large near-zone fields associated with plasmonic nanocolumns, do not preserve the polarization state due to the ellipsoidal cross sections of the columns and couple to the SPP waves on the two interfaces. The locations of the measured transmittance minima arising due to coupling with SPP waves are marked by arrows in  Fig.~\ref{transmission_spectra}.  The measured and simulated transmittance minima agree well.

The simulations clearly reveal that the coupling of the $s$-polarized light to the SPP waves can occur entirely due to the finite cross-sections of the nanocolumns. This validates the point that non-preservation of the polarization state in the scattered fields causes the coupling to  SPP waves. The indirect coupling of the incident $s$-polarized light excites SPP waves more symmetrically for $\theta\stackrel{>}{<}0$ than the direct coupling of the incident $p$-polarized light.  

However, we note that the overall transmittance is higher in the simulated results than the experimental measurements. This could arise due to various assumptions about the thickness of the metal layer (absent in the simulations) and the density of the columns along the grating lines. An additional mechanism for both the coupling of SPP waves to $s$-polarized light as well as the lowered transmittance can arise from imperfections in the fabricated structures which leads to slightly misaligned nanocolumns or bending of the nanocolumns along the grating direction. Such nanocolumns can create cross-polarization in the plane of incidence for $s$-polarized light which can also couple to SPP waves. 

\section{Concluding remarks}\label{conc}
We  fabricated periodically patterned CTFs of silver that can support the propagation of SPP waves on their interfaces with isotropic dielectric materials. These structures show strongly asymmetric diffraction and  can provide for unidirectional coupling to SPP waves which can be useful in many optical devices. The Bruggeman formalism was used to homogenize the silver CTFs into  homogeneous hyperbolic biaxial materials. The predicted  dispersions of the SPP waves matched the measured dispersions very well. The porosity can be controlled easily through the direction of metal-vapor flux during fabrication and is a critical parameter for controlling the dispersions. Both  the nanocolumnar tilt angle and length  need to be optimized for obtaining the desired  blaze action and  coupling strengths at the desired frequencies through full-wave 3-D photonic structure calculations. Our work creates a strong basis for the understanding of the photonic properties of plasmonic CTFs and devices based on such 
structured nanomaterials.

\begin{acknowledgments}
Funding by the Council of Scientific and Industrial Research (India) under Grant No. 03(1161)/10/EMR-II is acknowledged. J.D. thanks the University Grants Commission (India) for a fellowship and A.L. thanks the Charles Godfrey Binder Endowment at Penn State for ongoing financial support of his research.
\end{acknowledgments}

\end{document}